# Anomalous photoelectron spectrum caused by finite interaction time in few-cycle xuv laser pulses

Yang Xiang[1,2], Yueping Niu[1,*], Yihong Qi[1], and Shangqing Gong[1,**]

[1] State Key Laboratory of High Field Laser Physics, Shanghai Institute of Optics and Fine Mechanics, Chinese Academy of Science, Shanghai 201800, China

[2] College of Computer Science and Technology, Henan Polytechnic University, Jiaozuo 454000, China

*niuyp@siom.ac.cn, **sqgong@siom.ac.cn

**Abstract:** With the development of laser technology, pulse length enters the optical cycle regime and hence the interaction time between laser pulse and atoms becomes prominent. We investigate this problem in this Letter through the photoelectron spectrum of hydrogen atom in few-cycle xuv laser pulses. By solving one-dimensional time-dependent Schrödinger equation, we find that due to the insufficient interaction time, the electron can not gain enough energy from optical field when escaping the bind of the nuclear and then the abnormality appears in the photoelectron spectrum: the peak of photoelectron spectrum shows red shift compared with the well-known Einstein photo-electric effect formula. The shift becomes large as the pulse duration decreases.

When an atom absorbs a photon with energy larger than its bound potential, it will be ionized and a photoelectron will be generated. This process can be described by the well-known Einstein photo-electric formula, since the intensity of the light field was very low and the light duration was very long in those days when the photo-electric effect was found. Recently, due to the advanced laser technology, pulses with super-high intensity appear [1]. Therefore, photoionization of atoms in intense laser fields has attracted great interest around the world since the discovery of above-threshold ionization (ATI) by Agostini *el al.* [2]. The physical origin of ATI can be well explained by the simple-man model [3] and the Einstein photo-electric formula does not work here. ATI can be applied to the characterization of high-order harmonics and measurement of the absolute phase of few-cycle laser pulses [4]. Recently, many groups paid their attention to the ATI process in few-cycle pulses [5-7], since pulses in the near-infrared which are as short as a few optical cycles can be obtained in experiment [8].

On the other hand, with the development of the technology of high order harmonic generation, a single few-cycle xuv pulse with duration about 80 attosecond can be obtained [9]. As is known, attosecond pulse provides a power tool for the investigation of ultra-fast process on the molecular and electron level, such as the Auge-decay process [10], molecular dynamics [11], and the motion of the electronic wave packets in an atom [12]. In the few-cycle regime, we think that the interaction time between laser pulse and atoms shall become prominent. Hence, we will investigate this problem in this Letter through the photoelectron spectrum of hydrogen atom in few-cycle xuv laser pulses. To our knowledge, there are no detailed studies for the process of the photo-electric effect in few-cycle regime, such as how does the pulse duration impact on the PS and whether the Einstein photo-electric formula is still in effect, though many works have been done on the attosecond photoionization. For example, Yudin *et al.* found that attosecond PS and its asymmetry can monitor the coherent electron motion in atoms or molecules [13], Eckle *et al.* and Hentschel *et al.* used PS to measure the duration of attosecond pulse [14, 15], and Kienberger *et al.* pointed out that through the 'tomographic images' of the time-momentum distribution of the photoelectrons, one can look insight into the relaxation dynamics of the electronic shell following excitation [16]. Here, we consider a xuv pulse whose photon energy is so large that a single photon may deprive an electron from the atom and cause the ionization. On the opinion of quantum optics, the time that the atom spends on absorbing a photon is negligible. While in few-cycle regime, we show that the response time of photo-electric effect should be taken into account which has never been considered before.

We use one-dimension (1D) time-dependent Schrödinger equation (TDSE) to describe the interaction between the hydrogen atom and laser field in dipole approximation ($m=e=\hbar=1$, in atomic units, a.u.):

$$i\frac{\partial \psi(x,t)}{\partial t} = \left[ -\frac{1}{2}\frac{\partial^2}{\partial x^2} + V(x) - xE(t) \right] \psi(x,t), \quad (1)$$

where $V(x)$ is the Coulomb potential and $E(t)$ is the electric field of the attosecond laser pulse which can be obtained by $E(t) = -\partial A(t)/\partial t$, where $A(t)$ is the vector potential with the form of

$$A(t) = A_0 \sin^2\left(\frac{\pi t}{nT} + \frac{\pi}{2}\right) \cos(\omega t + \phi), \quad -\frac{1}{2}nT \leq t \leq \frac{1}{2}nT. \quad (2)$$

The parameters $A_0$, $\omega$, $\phi$ and $T$ are the amplitude, frequency, carrier envelope phase (CEP) and period of the laser pulse respectively. $n$ is the pulse length measured in optical cycle (OC). In all of our calculations, the peak intensity of the attosecond pulses is $1.0 \times 10^{15}$ W/cm$^2$, and $\omega$ is 2.278 a.u. (corresponding to wavelength $\lambda$ =20 nm). Equation (1) can be integrated using split-operator method [17]. When the time-dependent wave function $\psi(x,t)$ is obtained, we can then calculate the PS intensity by projecting the wave function onto the positive energy states,

$$P(W) = \left| \langle W | \psi(x,t) \rangle \right|^2, \quad (3)$$

where $W$ and $|W\rangle$ are the eigenvalues and eigenvectors of the unperturbed Hamiltonian of hydrogen atom. In order to smooth the PS, a linear interpolation for continuum energies is employed [18].

Now we present the results for the photoelectron spectra of the hydrogen atom in few-cycle xuv pulses. In the numerical simulations, the screened soft-core potential model $V(x) = -\alpha/\sqrt{1+x^2}$ is employed in order to avoid the unstable numerics caused by the Coulomb singularity [17, 19]. For 1D hydrogen atom, $\alpha$ is chosen as 0.775 a.u. so that the ionization potential $I_p$ of the model atom is 13.6 eV, which is equal to that of the real hydrogen atom [20]. Figure 1 shows the photoelectron spectra of 1D hydrogen atom in xuv pulses with different pulse lengths (the CEPs of these pulses are set to be zero). From this figure, one can see: with the decrease of the pulse length, (ⅰ) the intensity of the PS decreases dramatically, (ⅱ) the width of the spectrum becomes broad, and (iii) the peak photoelectron energy $E_p$ becomes red shift compared with the value $E_i = \omega - I_p$ (perpendicular line in Fig.1) which is obtained from the well-known Einstein photo-effect formula. It is easy to understand the cases of (ⅰ) and (ⅱ). The shorter the pulse is, the less the electrons can be deprived from atoms, which cause to the decrease of the PS intensity. Moreover, the bandwidth of the pulse becomes broad as the pulse length decreases and therefore the width of the PS becomes broad accordingly. For the case of (iii), it seems hard to understand that the peak photoelectron energy does not obey the Einstein photo-effect formula. While considering the response time of photo-electric effect, the abnormality in the PS may be understood. From the Fourier analysis, the duration of a photon with frequency of $\omega$ should be infinite. Therefore, it takes infinite time for an atom to absorb a photon theoretically. So, if the interaction time between the pulse and the atom is very short, then the electron can not gain enough energy (less than a photon) in the process of ionization. As a result, the peak photoelectron energy is less than $E_i$. The shorter the pulse length is, the less energy the photoelectron obtains when escaping the bind of the nuclear. Therefore, the difference between $E_p$ and $E_i$ increases as the pulse length shortens.

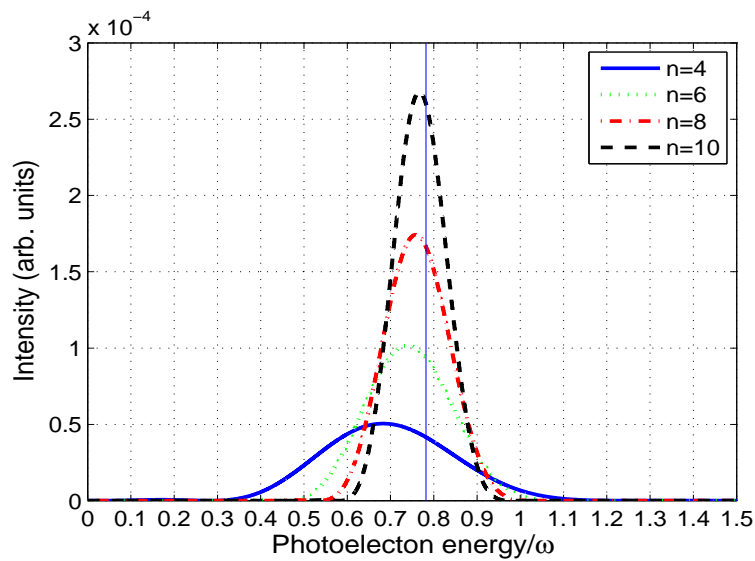

Fig.1. (Color online) The photoelectron spectra of 1D hydrogen atom in xuv attosecond pulses with different pulse lengths.

Based on the proposal mentioned above, we conclude that in few-cycle regime, the PS should not be sensitive to the CEP, since the PS dependents mainly on the interaction time, i.e. pulse length. To verify this, we calculate the PS in xuv attosecond pulses with different CEPs and the results are shown in Fig.2. Apparently, the PS laps over each other though the CEP varies from 0 to $\pi/2$, i.e. the CEP has no effects on the PS, just as we expect.

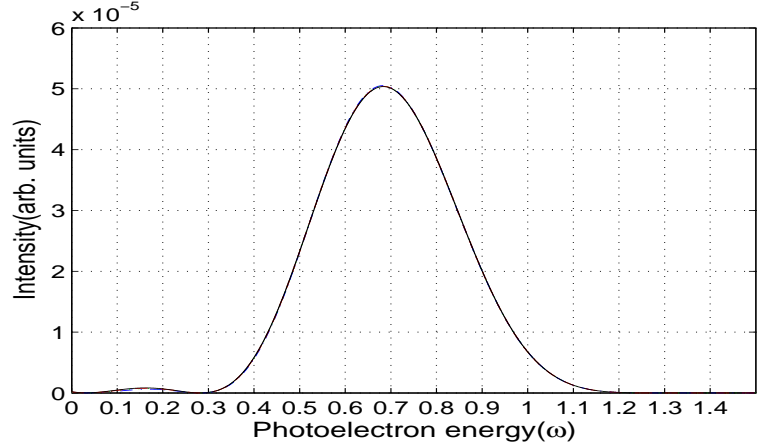

Fig.2.(Color online) The PS of 1D hydrogen atom in xuv attosecond pulses with different CEPs. The pulse length used is 4 OC.

Since the abnormality of the PS is caused by the insufficient time of the interaction between the laser pulse and the atom, the red shift should appear in the PS and have similar variation tendency with pulse length for any atoms and the pulse with any frequency, supposing that the frequency of the pulse is well above the ionization potential of the atoms. For 1D hydrogen and helium atom, the red shift $\Delta\,(=E_i-E_p)$ of the PS peak is shown in Fig.3 (a) and (b) respectively. In the calculation of the PS of 1D helium atom, single-active electron approximation [21] is used and the soft-core potential is $V(x)=-1/\sqrt{0.484+x^2}$ [22]. Clearly, similar tendency for each wavelength persists, i.e., the red shift $\Delta$ decreases as the pulse length increases.

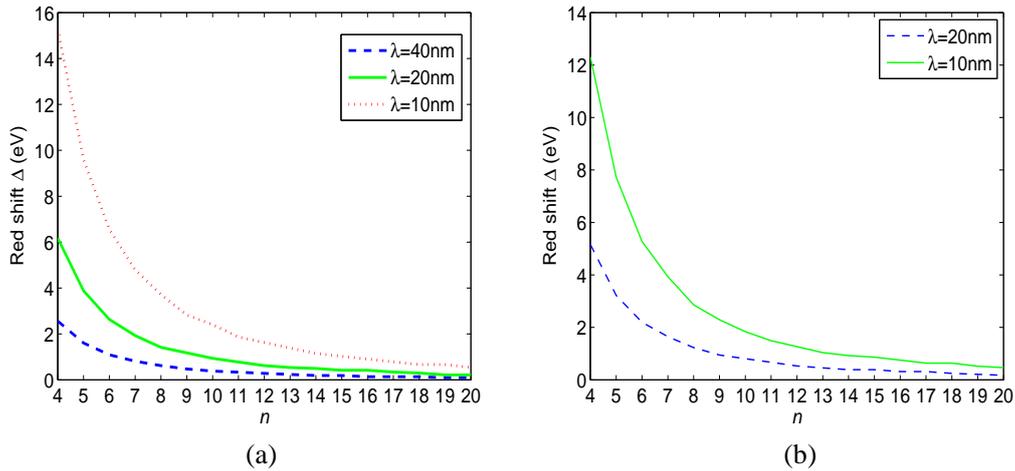

Fig.3. (Color online) Red shift of the PS peak of 1D hydrogen (a) and helium(b)atom as a function of pulse length and wavelength

In conclusion, we investigated the finite interaction time between laser pulse and atoms in few-cycle regime through the PS of hydrogen atom. Due to the insufficient time of the interaction between the atoms and the laser pulse, the electron can not gain enough energy from optical field when escaping the bind of the nuclear, which causes the abnormity in the PS, i.e., the peak of PS appears red shift compared with the well-known Einstein photo-electric effect formula.

**Acknowledgement**

The work is supported by the National Basic Research Program of China (Grant No.2006CB921104, 2006CB806000, 60708008, 10734080, 10874194, 60978013, and 60921004)